\documentclass[12pt]{article}
\usepackage{epsf, cite, amssymb}
\usepackage{epsfig}

\makeatletter
\renewcommand\section{\@startsection {section}{1}{\z@}%
                                   {-3.5ex \@plus -1ex \@minus -.2ex}
                                   {2.3ex \@plus.2ex}%
                                   {\normalfont\large\bfseries}}
\renewcommand\subsection{\@startsection{subsection}{2}{\z@}%
                                     {-3.25ex\@plus -1ex \@minus -.2ex}%
                                     {1.5ex \@plus .2ex}%
                                     {\normalfont\bfseries}}
\makeatother

\let\non\nonumber

\newcommand{\bea}{\begin{eqnarray}}
\newcommand{\eea}{\end{eqnarray}}
\newcommand{\be}{\begin{equation}}
\newcommand{\ee}{\end{equation}}

\newcommand{\R}{{\mathbb R}}

\newcommand{\rr}{\rightarrow}

\newcommand{\p}{\partial}

\def\slash#1{\setbox0=\hbox{$#1$}#1\hskip-\wd0\hbox to\wd0{\hss\sl/\/\hss}}
\newcommand{\mP}{\mathbb P}
\newcommand{\mC}{\mathbb C}

\newcommand{\D}[1]{\ensuremath{\mathrm{D}#1}}

\newcommand{\C}[1]{$(\ref{#1})$}

\begin{document}
\begin{titlepage}

\begin{center}

{March 7, 2003}
\hfill                  hep-th/0303066

\hfill EFI-03-09

\vskip 2 cm
{\Large \bf {World-sheet Stability of (0,2) Linear Sigma Models}}
\vskip 1.25 cm { Anirban Basu\footnote{email address:
basu@theory.uchicago.edu} and Savdeep Sethi\footnote{email address:
 sethi@theory.uchicago.edu}}\\
{\vskip 0.5cm  Enrico Fermi Institute, University of Chicago,
Chicago, IL
60637, USA\\}

\end{center}

\vskip 2 cm

\begin{abstract}
\baselineskip=18pt
We argue that   
two-dimensional $(0,2)$  gauged linear sigma models are not destabilized
by instanton
generated world-sheet superpotentials. We construct several examples where we show
this to be true. The general proof is based on the Konishi anomaly for
$(0,2)$ theories.
 
\end{abstract}

\end{titlepage}

\pagestyle{plain}
\baselineskip=19pt

\section{Introduction}
One of the basic questions we ask about quantum field theory is
whether the classical vacua are stable. Often, the structure of the
quantum moduli space is significantly different from the classical
moduli space. The theories about which we can usually say the most are
supersymmetric. In these cases, we can often make exact statements,
either perturbative or non-perturbative, 
because of non-renormalization
theorems. In cases where the vacuum structure is not renormalized at
any finite order in perturbation theory, non-perturbative effects, like
instantons, can still generate superpotentials which modify or destabilize
perturbative vacua. Numerous examples of this kind have been studied
in various dimensions; for example, N=1 supersymmetric QCD in four  
dimensions~\cite{Affleck:1984mk}. 

The aim of this work is to study the stability of 
two-dimensional gauge theories, both massive and massless, with
$(0,2)$ world-sheet supersymmetry. On the string world-sheet, the
terminology $(p,q)$ supersymmetry
refers to theories with $p$ left-moving and $q$ right-moving
supersymmetries. Conformal field theories with
$(0,2)$ supersymmetry are a key ingredient in building perturbative
heterotic string compactifications (for a review,
see~\cite{Distler:1995mi}). 
Unlike their $(2,2)$ cousins,
theories with $(0,2)$ supersymmetry that are conformal to
all orders in perturbation theory can still be destabilized by
world-sheet instantons. The usual phrasing of this problem is that
world-sheet instantons generate a {\it space-time} 
superpotential~\cite{Dine:1986zy, Dine:1987bq}. However, the general belief is
that this destabilization is generic. Under special conditions
described in~\cite{Distler:1987wm, Distler:1988ee}\ for non-linear sigma models, there
can be extra fermion zero
modes in an instanton background which kill any non-perturbative
superpotential. 

We consider those $(0,2)$ models which can be constructed as IR limits
of gauged linear sigma models~\cite{Witten:1993yc}. This is a rather
nice class of models which can be conformal or non-conformal, and
which can flow to theories with IR descriptions like sigma models or
Landau-Ginzburg theories. For perturbatively conformal cases, some
criteria for the absence of a space-time superpotential have been
described in~\cite{Silverstein:1995re, Distler:1994hq,
  Distler:1994hs}. 
Our interest is in whether a
world-sheet superpotential is generated. In perturbatively conformal
cases, the two questions should be related in a way that we will
describe. 

In the following section, we
consider the stability of $(0,2)$ theories without tree level
superpotentials. We construct several examples of non-conformal
$(0,2)$ models without tree level superpotentials for which we show
that no world-sheet superpotential is generated by instantons. This
result surprised us initially since we were looking for a model with
an instanton generated superpotential! In section three, we give a
general  argument based on the Konishi
anomaly~\cite{Konishi:1984hf,Konishi:1985tu} that this is true for all
gauged linear sigma models without tree level superpotentials. This
argument is inspired in part by recent progress in four-dimensional
gauge theories~\cite{Dijkgraaf:2002dh, Cachazo:2002ry}. We then
extend  the argument to cases with a tree level
superpotential. In all cases, it appears that a non-perturbative
superpotential is forbidden. 

Lastly, we consider the implication of our results for the
space-time superpotential. Based on the absence of a non-perturbative
world-sheet superpotential,  we argue that there is no corresponding 
space-time instability. Some related observations will appear
in~\cite{toappear}.

\section{Some (0,2) Examples}

In this section we construct examples of $(0,2)$ gauged linear 
sigma models~\cite{Witten:1993yc}\  without tree level
superpotentials. We will show that no superpotential is generated by
non-perturbative instanton or anti-instanton effects. It is actually sufficient to
consider one instanton contributions. Higher instanton numbers
generate more fermion zero modes which obstruct the generation of a
superpotential. 

\subsection{A bundle over $\mC\mP^3$}

The $(0,2)$ superspace and superfield notations are reviewed in
Appendix \ref{superspace}. We begin by considering a $U(1)$ gauge theory.
The $(0,2)$ action is given by a sum of terms
\be S=S_g + S_{ch} + S_F + S_{D \theta} + S_J\ee
where $S_g, S_{ch}, S_F$ are canonical kinetic terms for the
gauge-fields, bosonic chiral superfields, and fermionic chiral
superfields, respectively. The explicit form of these actions appear
in Appendix \ref{superspace}. The Fayet-Iliopoulos $D$-term and  
the theta angle appear in $S_{D \theta}$ while $S_J$ contains any tree
level superpotential. For these models, we set $S_J=0$. 

As our first example, we construct a linear sigma model whose IR
limit is a non-linear sigma model on $C\mP^3$. This is a cousin of the
$(2,2)$ model studied in~\cite{Penin:1998iv}. 
Apart from the $U(1)$ gauge superfields $\Psi$ and $V$,  we have
bosonic superfields $\Phi_i =\phi_i + \ldots$ where $i=1,\ldots, 4$, and a single Fermi
superfield $\Gamma$.  Each $\Phi_i$ carries gauge charge
$1$ while $\Gamma$ carries gauge charge $-2$. 
Since we do not have a tree level superpotential our action is
\be S=S_g + S_{ch} + S_F + S_{D \theta}. \ee
Solving for the auxiliary fields gives the following bosonic potential for the $\phi_i$
 \be U= \frac{D^2}{2 e^2}= \frac{e^2}{2} ( \sum_{i} \vert \phi_i\vert^2-r)^2.\ee 
We take $r$ to be positive, and set $r= \eta^2$. After modding by the
$U(1)$ gauge symmetry, we see that the target space is $C\mP^3$.  

The Fermi superfield determines the gauge-bundle over  $C\mP^3$ which,
in this case, is the line bundle ${\cal{O}}(-2)$. 
These particular gauge charge assignments guarantee gauge anomaly
cancellation, which is a basic consistency requirement. 
This can be seen either by computing the requisite one loop diagrams,
or by checking that the condition for anomaly cancellation~\cite{Distler:1987wm}
\be {\rm{ch}}_2 (TM) = {\rm{ch}}_2 (V) \ee
is satisfied. Here, $TM$ is the tangent bundle of $C\mP^3$, $V$
is the ${\cal{O}}(-2)$ line bundle, and ${\rm{ch}}_2$ is the second Chern character. 
Using the definition 
$${\rm{ch}}_2 (X) = \frac{1}{2} {\rm{c}}_1^2 (X) - {\rm{c}}_2 (X)$$ we
see that both sides of this equation gives $2 J^2$, where $J$ is the
curvature $2$-form of the hyperplane bundle over $C\mP^3$.

This theory is massive (like the $(2,2)$ $C\mP^3$ model) because the
sum of the gauge charges of the
right moving fermions is non-zero. The theory does, at the classical level, have a chiral
$U(1)$ symmetry under which $(\psi_+^i,\lambda_-,\chi_-)$ carry
charges ($1,-1,q$), where $q$ is any integer.  This symmetry is
anomalous at one-loop for any  $q \neq -2$. The charge of the gaugino,
$\lambda_-$, does not matter in the anomaly computation because it is
not charged in a $U(1)$ theory. Since this chiral symmetry is
anomalous, we can shift the theta angle to any value, and we choose to
set it to zero.

Let us now consider how instantons modify the perturbative theory.
First we construct the one instanton BPS solution. In order to construct
an instanton solution, we wick rotate to Euclidean space sending 
$$y^0 \rr  -i y^2, \quad v_{01} \rr -iv_{12}.$$ 
The Euclideanised bosonic action is
\be \label{eact} S= \int {\rm{d^2}} y \ [ \ \frac{1}{2e^2} v_{12}^2 + \sum_{i}
\vert  \D_{\alpha}\phi_i\vert^2 +\frac{e^2}{2}( \sum_{i} \vert \phi_i\vert^2-\eta^2)^2 ]. \ee
We construct the well known vortex instanton solution~\cite{vortex,
  Nielsen:1973cs}\ of the Abelian Higgs model in two dimensions: take
$\phi_i$=0 for $i$=2,3,4 and take non-zero $\phi_1$ and gauge
fields. From now on  we refer to $\Phi_1$ as $\Phi$ for brevity. We
also set $e=1$. In polar coordinates, the one-instanton configuration is given by
\be v_r =0, \quad  v_{\theta}= v(r), \quad   \phi = f(r) e^{i \theta}\ee
where for large $r$, 
\be v(r) \sim \frac{1}{r} + {\rm{constant}} \times  \frac{e^{- \eta r}}{\sqrt r},\ee
\be f(r) \sim \eta + {\rm{constant}} \times  e^{-{\sqrt 2} \eta r},\ee
and $v(0)=f(0)=0$. 
The Bogomolnyi equations are
\be \label{bone}(D_1 +i D_2)\phi=0\ee
and 
\be \label{btwo} D + v_{12}=0.\ee
On evaluating \C{eact}\ in this background, we easily obtain the usual
instanton  action $S=2 \pi \eta^2$. Next, we
are interested in  constructing the fermion zero modes in this
instanton background.  They are explicitly given by
\be \label{fzm}\mu^0 = \pmatrix{ {\bar\psi}_+^0 \cr \lambda_-^0\cr} = \pmatrix{
  -{\sqrt 2} ({\bar{D}}_1  + i{\bar{D}}_2){\bar\phi} \cr D-v_{12}\cr}\ee 
\be{\bar\chi}_-^0  =\phi^2, \ee
and
\be {\bar{\psi}}_{+i}^0 = \bar{\phi}  \ee
for $i=2,3,4$. Note that these zero modes are normalizable because of
the exponential fall off of the fields at large distances. The $\mu^0$ fermion
zero mode is actually the zero mode generated by the broken
supersymmetry generator. In
order to see this, we must examine the supersymmetry transformations
in the instanton background. 

The supersymmetry transformations become involved because we must also
make gauge transformations to preserve Wess-Zumino gauge. The relevent
supersymmetry transformations are given by
\bea \delta {\bar{\psi}}_+ &=& -i {\sqrt 2} ({\bar{D}}_0 + {\bar{D}}_1)
\bar{\phi} \epsilon_-, \cr
 \delta \lambda_- &=& i D \epsilon_- + v_{01} \epsilon_-. \eea
The supersymmetry parameter, $\epsilon_-$, corresponds to $Q_+$. Wick
rotating to Euclidean space gives the zero mode found in
\C{fzm}. Hence, $Q_+$ is the broken supersymmetry while the
${\bar{Q}}_+$ supersymmetry is still preserved by the instanton
background. Using the Bogomolnyi equations, it is not hard to check
that the $\mu^0$  zero mode does satisfy $i{\slash{D}} \mu^0 =0$
where $i {\slash{D}}$ is the Dirac-Higgs operator
\be i {\slash{D}} = \pmatrix{ -i ( {\bar{D}}_1-i {\bar{D}}_2) 
& {\sqrt 2} i \bar{\phi} \cr -{\sqrt 2} i \phi & i \partial_1
  -\partial_2\cr}. \ee
The field ${\bar\chi}_-$ is expanded in modes of the Dirac operator 
$({\bar{D}}_1+ i {\bar{D}}_2)$ (note that $\Gamma$ has gauge charge
$-2$). The ${\bar{\psi}}_{+i}$ ($i=2,3,4$) fields are expanded in modes
of the Dirac operator $({\bar{D}}_1 -i {\bar{D}}_2 )$. We ask again
whether there are any zero modes for these fields. 
The existence of zero modes for these operators can be predicted using
 index theory and a vanishing theorem~\cite{Nielsen:1977aw}. Note that 
in Minkowski space, ${\bar{\psi}} = \psi^{\dagger} \gamma^0 = 
({\bar{\psi}}_+ ~{\bar{\psi}}_-)$. However, in Euclidean space 
$\psi$ and ${\bar{\psi}}$ are independent fermionic fields and not 
the conjugates of each other. Here, ${\bar{\psi}}= (\eta_- ~\eta_+)$ 
and so in the Euclidean formulation of the theory, the 
$\mu, {\bar{\psi}}_{+i}$ ($i=2,3,4$) zero modes are zero modes of negative 
chirality while the ${\bar{\chi}}_-$ zero mode is a zero mode of 
positive chirality.

We can now ask what gauge invariant correlators are non-vanishing in
this instanton background. There are only two
possibilities
\be \langle {\bar{\psi}}_+ {\bar{\psi}}_{+2} {\bar{\psi}}_{+3}
    {\bar{\psi}}_{+4} {\bar{\chi}}_- \phi^2 \rangle, \qquad  \langle
    \lambda_- {\bar{\psi}}_{+2}{\bar{\psi}}_{+3}{\bar{\psi}}_{+4}
	   {\bar{\chi}}_- \phi \rangle. \ee  
which can have a non-zero vacuum expectation value in the instanton
background. However, neither of these terms could be generated by a
term in the $(0,2)$ superpotential since there are far too many
fermion zero modes. A superpotential term could absorb, at most, two
fermion zero modes. Therefore, we see that there is no
instanton generated superpotential. The same argument applies to
instantons embedded in the other $\phi_i$.

 Next we show that there is no superpotential generated by a one
 anti-instanton  contribution. The details are very similar to the one
 instanton case so we shall be brief. The ani-instanton configuration
 is similar to the instanton case except
\be \phi = f(r) e^{-i \theta}\ee
and for large $r$, 
\be v(r) \sim -\frac{1}{r} + {\rm{constant}} \  \frac{e^{- \eta
 r}}{\sqrt r}.\ee   The Bogomolnyi equations are now
\be (D_1 -i D_2)\phi=0\ee
and 
\be D - v_{12}=0\ee
leading to the anti-instanton action $S=2 \pi \eta^2$. The
 normalizable fermion zero  modes are now given by
\be \mu^0 = \pmatrix{ \psi_+^0 \cr {\bar{\lambda}}_-^0\cr} =  
\pmatrix{ -{\sqrt 2} (D_1 + i D_2)\phi \cr D+ v_{12}\cr}\ee 
\be \chi_-^0  ={\bar{\phi}}^2\ee
and
\be {\psi}_{+i}^0 = \phi  \ee
for $i=2,3,4$. Again the fermion zero mode analysis rules out the
 generation of a gauge invariant superpotential. Hence we see that
 there is no superpotential generated by a one anti-instanton
 contribution.

\subsection{Changing the bundle}

 We can also consider the case where $\Gamma$ carries gauge charge
 $2$. This leads to a ${\cal{O}}(2)$ line bundle over $C\mP^3$. The
 gauge anomaly cancellation condition is satisfied as the anomaly is
 proportional to the square of the charges. The model is still
 massive, and we again ask whether a superpotential is generated. 

In this case, we find the same $\mu^0, {\bar{\psi}}_{+i}^0$ zero modes
(for $i=2,3,4$) in the instanton background. However, there is now a zero
mode for the field $\chi_-$ which is expanded in modes of the Dirac
operator $(D_1 + i D_2)$. The zero mode is given by
$$\chi_-^0= \phi^2. $$
Again this is normalizable given the exponential decay of the fields
at large distances. Once again a gauge invariant superpotential cannot
be generated. Similar arguments hold for the case of the
anti-instanton. Hence for both line bundles, ${\cal{O}} (\pm 2)$, over
$C\mP^3$,  no world-sheet superpotential is generated. These theories
are non-perturbatively stable.

\subsection{A different route to stability}

In the previous examples, a superpotential was forbidden because of
the large number of zero modes in an (anti-)instanton background. 
We now turn to an example where we have the right number of fermion
zero modes for a superpotential, but we will show that even in this
case, there is no superpotential generated. 

We consider a theory with one bosonic superfield, $\Phi$, carrying
gauge charge $1$ and one Fermi superfield, $\Gamma$, carrying gauge
charge $-1$. There are also the required gauge superfields $\Psi$ and
$V$. This gauge charge assignment causes the gauge anomaly to
cancel. We can also set the theta angle to zero because of the
non-zero chiral anomaly. We again construct vortex instanton solutions
satisfying the Bogomolnyi equations \C{bone}\ and \C{btwo}. The
fermion zero modes are $\mu^0$ as in \C{fzm}\ and ${\bar{\chi}}^0_- =
\phi$. In this case, we see that 
$$\langle {\bar{\psi}}_+ {\bar{\chi}}_- \rangle$$ 
can get a non-zero vacuum expectation value in the instanton
background. This would lead to the existence of a superpotential
\be S= -\frac{a}{\sqrt 2} \ \int {\rm{d^2}} y \ {\rm{d}}
    {\bar{\theta}}^+\  \bar{\Gamma} \bar{\Phi} \vert_{\theta^+ =0} \ee
where $a$ is a constant that can be determined. Hence the $(0,2)$
theory would be rendered unstable by this non-perturbative
effect. However, the condensate has a vacuum expectation value proportional to
\be \label{zint}
\int {\rm{d^2}} x_0 \ \phi ({\bar{D}}_1 +i {\bar{D}}_2) \bar{\phi}\ee
where we have integrated over the two bosonic translational zero
modes~\cite{Gervais:1975dc, Tomboulis:1975gf}. Using the identity
\be 2i \phi ({\bar{D}}_1 +i {\bar{D}}_2) \bar{\phi} + (\partial_1 +i
\partial_2 )  (D - v_{12}) =0,\ee
which can be proven using the Bogomolnyi equations, we see that this
integral is actually zero!  Yet again there is no instanton generated
superpotential.

Lastly, we consider the case where $\Gamma$ has a gauge charge $1$. In
 this case,  we can obtain $\Phi$ and $\Gamma$ from a single $(2,2)$
 chiral superfield. The fermion zero modes are $\mu^0$ as in \C{fzm}\
 and $\chi_-^0 = \phi$. However, the possibility $\langle
 {\bar{\psi}}_+ \chi_- \rangle$ cannot be generated from a
 superpotential bacause of holomorphy of the
 superpotential. Therefore, no superpotential is generated at
 all. This certainly agrees with the $(2,2)$ non-renormalization
 theorem. From these examples, we see no non-perturbative
 superpotential generated. These results together with our other
 attempts at finding  examples with
 non-vanishing superpotentials suggest that this phenomena is quite
 generic. In the next section, we give a general argument explaining why this happens.   
    
\section{The $(0,2)$ Konishi Anomaly}
\subsection{Deriving the anomaly}

  In this section, we obtain the Konishi anomaly~\cite{Konishi:1984hf,Konishi:1985tu}\
for the $(0,2)$ linear sigma model with no tree level
superpotential. Because of a perturbative non-renormalization theorem,
the only superpotential that can possibly be generated is a
non-perturbative one. From the Konishi anomaly relation that we
obtain, we argue that no such non-perturbative superpotential can be
generated  by instantons. This generalizes the results of the previous
section.

 Our derivation of the Konishi anomaly is along the lines
 of~\cite{Konishi:1985tu}\ which is a superspace generalization of
 Fujikawa's functional integral method~\cite{Fujikawa:1979ay}. 
We start with the linear sigma model with no tree level
 superpotential. We assume that a superpotential is generated
 non-perturbatively. Hence, 
$$S = S_g + S_{ch} + S_F + S_{D \theta} +
 S_J$$ 
where $S_J$ is the non-perturbative superpotential
 contribution. We want to prove that $S_J=0$ in the (anti-)instanton
 background. We denote all the chiral superfields by $\Sigma$, i.e.,
 $\Sigma = \{\ \Phi_i^0, \Gamma_a^0 \}\ $, and the corresponding
 anti-chiral superfields by $\bar{\Sigma}$. The partition function is
 given by 
\be Z= \int[D {\Phi_i}^0 D {\bar{\Phi_i}}^0 D {\Gamma}_a^0 D  
{\bar{\Gamma}}_a^0 D \Psi D V ] \ e^{iS}. \ee
In the functional integral formalism, all the fields in the path
 integral measure, 
$$\Phi_i^0, {\bar{\Phi}}_i^0, \Gamma_a^0, {\bar{\Gamma}}_a^0, \Psi,
 V,$$ 
are independent, and one can study their transformations separately. 

We consider a global axial $U(1)$ transformation  given by
$$\Sigma_m \rightarrow e^{i\alpha}\Sigma_m.$$ 
The subscript $m$ signifies that only one of the fields in $\Sigma$
transforms non-trivially. To extract a Ward identity in superspace, we
consider the following transformation
\be \Sigma_m \rightarrow \Sigma_m'= e^{i A} \Sigma_m \ee
where $A$ is a chiral superfield satisfying ${\bar{D}}_+ A =0$. This
leads to a change in the measure and action
\be Z= \int [D \Sigma_m ...] \ e^{i S}= \int [D \Sigma_m' ...] \  
e^{i S'}= \int [D \Sigma_m ...] \ {\cal{J}}\ e^{i S + i \delta S}. \ee
{}For an infinitesimal transformation,
\be\delta S= -\frac{i}{2} \int {\rm{d^2}} y \ {\rm{d^2}} \theta   
{\bar{\Phi}}_i ({\cal{D}}_0 -{\cal{D}}_1) i A \Phi_i -\frac{1}{\sqrt
  2} 
\int {\rm{d^2}} y \ {\rm{d}} \theta^+ \sum_{a} \Gamma_a \frac{\delta
  J^a}{\delta \Phi_i}  i A \Phi_i \vert_{{\bar{\theta}}^+=0} \ee     
if $\Sigma_m$ is a bosonic chiral superfield, and
\be\delta S= -\frac{1}{2} \int {\rm{d^2}} y \ {\rm{d^2}} \theta
	  {\bar{\Gamma}}_a   i A \Gamma_a -\frac{1}{\sqrt 2} \int
	  {\rm{d^2}} y \  {\rm{d}} \theta^+ i A \Gamma_a  J^a  \vert_{{\bar{\theta}}^+=0} \ee 
if $\Sigma_m$ is a Fermi superfield. Also,
\be \label{jac} {\cal{J}}= {\rm{det_{c}}} (\frac{\delta \Sigma_m '}{\delta
  \Sigma_m})= {\rm{det_{c}}} (-i A {\bar{D}}_+)= e^{{\rm{tr_{c}}} (- i A {\bar{D}}_+)}.\ee
The reason for the subscript $c$, which means chiral, will be clear in
a moment. The trace originally involves integration over $y$ and
$\theta^+, \bar{\theta}^+$. Using the relation
\be \label{simplify}\int {\rm{d}} {\bar{\theta}}^+= \frac{\partial}{\partial
  {\bar{\theta}}^+}  = - {\bar{D}}_+ +i \theta^+ (\partial_0 +
  \partial_1), \ee
we have replaced the integral over  $\bar{\theta}^+$ by an insertion
  of $ - {\bar{D}}_+$ in \C{jac}. The second term in \C{simplify}\ is
  a total derivative which we can drop since we integrate over $y$. 
The remaining superspace integral in the chiral trace only involves
$\int {\rm{d}}^2 y \, {\rm{d}}\theta^+$, and is therefore a chiral integral.

We regulate the trace in the following way
\be \label{regtr} {\rm{tr_{c}}}^{{\rm{reg}}} (- i A {\bar{D}}_+)=  {\rm{lim}}_{M
  \rightarrow \infty} {\rm{tr_{c}}} (- i A  e^{\frac{L}{M^2}} {\bar{D}}_+) \ee   
where 
\be L= -\frac{i}{2} {\bar{D}}_+ \  e^{-\Psi} ({\cal{D}}_0 -
    {\cal{D}}_1) \  e^{-\Psi} D_+  \  e^{2 \Psi}. \ee   
Note that $L$ respects manifest supersymmetry and is chiral because ${\bar{D}}_+ L=0$. 
The $U(1)$ gauge transformation acts by
\be e^{\Psi} \rightarrow e^{-i\bar{\Lambda}} e^{\Psi} e^{i\Lambda},
\qquad  V \rightarrow  V + (\partial_0 -\partial_1)({\bar{\Lambda}}+\Lambda). \ee 
So under a gauge transformation
\be L \rightarrow L'= e^{-2i \Lambda } L e^{2i \Lambda}. \ee
Hence $L$ is gauge covariant as well. We now proceed to compute the
regulated trace in \C{regtr}. We have 
\be \label{exL} L= -\frac{i}{2} e^{-\Psi} \Upsilon e^{-\Psi} D_+ \ e^{2 \Psi}
-\frac{i}{2} e^{-\Psi} ({\cal{D}}_0 -{\cal{D}}_1) (2i({\cal{D}}_0 +
{\cal{D}}_1)- {\cal{D}}_+ {\bar{{\cal{D}}}}_+ ) e^{\Psi}. \ee 
{}From the regulated trace in \C{regtr}, it is clear that $L$ always
acts on ${\bar{D}}_+$.  So we have a non zero contribution only if we
have a factor of $D_+$ along with ${\bar{D}}_+$ since 
$$\langle D_+ {\bar{D}}_+\rangle =-1.$$ 
This is possible when one factor of $\Upsilon D_+$ is brought down
from the exponential. To get a non-zero contribution, we have to set
$\Psi=0$ in the first term in the expression for $L$. The last term in
\C{exL}\ involving  ${\cal{D}}_+ {\bar{{\cal{D}}}}_+$ does not
contribute. Also the second term involving $({\cal{D}}_0 +
{\cal{D}}_1)$ term contributes with  $\Psi=0$. So acting on ${\bar{D}}_+$,
\be \label{simL}  L= -\frac{i}{2} \Upsilon D_+  +({\cal{D}}_0 -{\cal{D}}_1)
({\cal{D}}_0 +{\cal{D}}_1).\ee
The leading term in the regulated trace is given by dropping the
background  gauge field terms in the second term in \C{simL}\ leading to
\be L= -\frac{i}{2} \Upsilon D_+ +({{\partial}_0}^2 -{{\partial}_1}^2).\ee 
Hence, the regulated trace gives
\be {\rm{tr_{c}}}^{{\rm{reg}}} (-i A {\bar{D}}_+)= i \int {\rm{d^2}} y
\ {\rm{d}} \theta^+  \frac{\Upsilon A}{8 \pi}.\ee   
Finally, we obtain the Ward identity
\be \frac{1}{2} {\bar{D}}_+ {\bar{\Phi}}_i ({\cal{D}}_0 - {\cal{D}}_1)
\Phi_i  = -\frac{i}{\sqrt 2} \sum_{a} \Gamma_a \frac{\delta
  J^a}{\delta \Phi_i}  \Phi_i \vert_{{\bar{\theta}}^+ =0} +\frac{\Upsilon}{8 \pi} \ee
for the bosonic chiral superfields, and
\be \label{detward} \frac{1}{2} {\bar{D}}_+ {\bar{\Gamma}}_a \Gamma_a = \frac{1}{\sqrt
  2}  \Gamma_a J^a \vert_{{\bar{\theta}}^+ =0} + i \frac{\Upsilon}{8 \pi} \ee
for the Fermi superfields. They can be combined and written as
\be \label{ward} {\bar{D}}_+ J = i \frac{\delta S_J}{\delta \Sigma_m}
\Sigma_m  \vert_{{\bar{\theta}}^+ =0}  + \frac{\Upsilon}{8 \pi}. \ee
Equation \C{ward}\ and its conjugate obtained from considering
anti-chiral transformations are the Konishi anomaly equations for the
$(0,2)$ linear  sigma model. 

\subsection{Applying the Konishi equations}

Now the relation \C{ward}\ is a rather beautiful operator relation. We
can take the expectation value of \C{ward}\ in a BPS (anti-)instanton
background. The left hand side is trivial in the chiral ring, and
vanishes by fermion zero mode counting.  We therefore obtain the general result
\be  i \langle \frac{\delta S_J}{\delta \Sigma_m }\Sigma_m
\vert_{{\bar{\theta}}^+  =0} \rangle = - \langle \frac{\Upsilon}{8 \pi} \rangle\ee
for all $m$. Similarly from anti-chiral transformations, we obtain
\be i \langle \frac{\delta S_J}{\delta {\bar{\Sigma}}_m }
    {\bar{\Sigma}}_m  \vert_{\theta^+ =0} \rangle = \langle
    \frac{\bar{\Upsilon}}{8 \pi}  \rangle \ee
for all $m$.  From
the component  expansion for $\Upsilon$ and $\bar{\Upsilon}$, we see
that the lowest component and the top component have vanishing vacuum
expectation value because of Lorentz invariance: they involve the one
point function of a fermion. The middle component of $\Upsilon$ is 
$$2i \theta^+ (D -i v_{01})$$ 
while that of $\bar{\Upsilon}$ is 
$$-2i {\bar{\theta}}^+ (D +i v_{01}).$$ 
Wick rotating to Euclidean space, we find that
\be \label{firstkon} \langle \Gamma_a J^a \vert_{{\bar{\theta}}^+ =0} \rangle = 
\frac{\theta^+}{2 \sqrt 2 \pi} \langle D - v_{12}\rangle \ee
and
\be \label{secondkon}\langle {\bar{\Gamma}}_a {\bar{J}}^a \vert_{\theta^+ =0} 
\rangle = \frac{{\bar{\theta}}^+ }{2 \sqrt 2 \pi} \langle D + v_{12} \rangle \ee
for all $a$. In the (anti-)instanton background, both $\langle D
\rangle$ and $  \langle v_{12} \rangle$ vanish because of fermion zero
modes.  Actually for theories with a broken chiral $U(1)$ symmetry,
this vanishing also follows independently from the Bogomolnyi equations. For
example, in an instanton background  we
see that \C{secondkon}\ vanishes using the Bogomolnyi equation \C{btwo}. The
right hand side of \C{firstkon}\  is proportional to $\langle v_{12}
\rangle$ which, in turn, is proportional to
$\theta$~\cite{Coleman:1976uz}.  However, because of the broken chiral
$U(1)$ symmetry, all theta vacuua are equivalent and we can set theta to
zero. The same analysis holds for the anti-instanton case leading to
the final result that  in an (anti-)instanton background 
\be \Gamma_a J^a \vert_{{\bar{\theta}}^+ =0}  =  {\bar{\Gamma}}_a  
{\bar{J}}^a \vert_{\theta^+ =0} =0. \ee
We  conclude that $S_J=0$, and no non-perturbative
superpotential is generated. 

\subsection{Cases with tree level superpotentials}

What changes when we add a tree level superpotential? It appears that
not a great deal changes in the preceeding argument. 
We replace $S_J$ by the sum of the tree
level superpotential, $S_J^{0}$, and any non-perturbative
superpotential,  $S_J^{non}$. The derivation just given goes through
without further change, and we obtain the same equations
\C{firstkon}\ and \C{secondkon}. Evaluated in an instanton background,
it again appears that the total superpotential must vanish. At first sight,
this might appear to be a contradiction since, by construction,
$S_J^0$ is non-zero. 

However, the condition for an instanton to be BPS is now modified. In
the presence of a superpotential, the BPS condition requires~\cite{Witten:1993yc}
\be
J_a^0 = 0 
\ee
so the Konishi relation is satisfied. Beyond multiplicatively
renormalizing $S_J^0$, it seems that a  non-perturbative
superpotential is again ruled out. 

\subsection{The space-time superpotential}

Lastly, for perturbatively conformal models, we want to address the
question of whether the absence of a
world-sheet superpotential implies the absence of a space-time
superpotential. In models with no tree level superpotential, we can
argue this relation as follows: a space-time superpotential implies
that our perturbatively conformal theories, which we can label by the
parameter $t = ir + \theta/2\pi$, do not flow to a family of
superconformal field theories
with a corresponding $t$ modulus. Let us just consider the dependence
on $r$. For example, they might flow to a
trivial theory with $r\rr \infty$.  

How can $r$ be renormalized? In the action, $r$ appears in the
term
$$ -r \int {\rm{d^2}} y \ D. $$
We need to ask whether $D$ can be renormalized in an instanton
background. Now $D$ is bosonic, and we must absorb fermion zero
modes. Where can they come from? The only place we see is a
non-perturbative superpotential. The perturbative Lagrangian will not
do because the zero modes are chiral. Since no world-sheet superpotential is
generated, $r$ remains an exactly marginal parameter and no
space-time superpotential is generated. 

What if there is a tree level world-sheet superpotential? In this case, fermion
zero modes could be absorbed from the Yukawa terms generated from the
superpotential. So fermion zero mode counting does not rule out
renormalization of $r$. 
However, using the Bogomolnyi equations, the remaining bosonic
integral is always of the form
$$ \int d^2 x_0  \, |\phi|^k (\p_1 + i \p_2) |\phi|^2 $$
where $k$ is a non-negative integer, and we have embedded the
instanton in $\phi$.  However, this integral
over the two
translational zero modes (with $\R^2$ as the Euclidean world-sheet)
vanishes. Again, it appears that no
space-time superpotential is generated.

\section*{Acknowledgements}
A.~B. would like to thank J.~Harvey for many useful discussions, and
A.~Penin for useful email correspondence. We would also like to thank
J.~Distler for extensive discussions and D.~Tong for helpful correspondence.
The work of A.~B. is supported in part by  NSF Grant No.
PHY-0204608.
The work of S.~S. is supported in part by NSF CAREER Grant No.
PHY-0094328, and by the Alfred P. Sloan Foundation.

\vfill\eject
\appendix
\section{The Structure of $(0,2)$ Superspace}
\label{superspace}

We review $(0,2)$ superspace following~\cite{Witten:1993yc}. 
We shall be dealing with abelian gauge theories. The superspace for
$(0,2)$ theories has 
bosonic coordinates $y^0, y^1$ and fermionic coordinates $\theta^+,
{\bar{\theta}}^+$. 
The supersymmetry generators act in superspace in the following way 
\bea Q_+ &=& \frac{\partial}{\partial \theta^+} +i {\bar{\theta}}^+
(\partial_0 + \partial_1), \\
{\bar{Q}}_+ &=& -\frac{\partial}{\partial {\bar{\theta}}^+} -i
\theta^+ (\partial_0 + \partial_1). \eea
On the other hand, the superspace covariant derivatives are given by
\bea D_+ &=& \frac{\partial}{\partial \theta^+} -i {\bar{\theta}}^+
(\partial_0 + \partial_1), \\
 {\bar{D}}_+ &=& -\frac{\partial}{\partial {\bar{\theta}}^+} +i
\theta^+ (\partial_0 + \partial_1). \eea
The following multiplets and the corresponding actions are used in
various sections of the main text.

\subsection{The gauge multiplet}

The superspace gauge covariant derivatives ${\cal{D}}_+,
{\bar{\cal{D}}}_+$ and 
$\cal{D_{\alpha}}$ ($\alpha=1,2$) satisfy the algebra
\be {\cal{D}}_+^2 = {\bar{\cal{D}}}_+^2 =0, \{\ {\cal{D}}_+,
    {\bar{\cal{D}}}_+ \}\ =2i ({\cal{D}}_0+{\cal{D}}_1). \ee
The first two equations imply that ${\cal{D}}_+ = e^{-\Psi} D_+
e^{\Psi}$ and ${\bar{\cal{D}}}_+ = e^{\bar{\Psi}} {\bar{D}}_+ e^{-\bar{\Psi}}$
where $\Psi$ takes values in the Lie algebra of the gauge group. In  Wess-Zumino gauge, 
$$\Psi = \theta^+ {\bar{\theta}}^+ (v_0 + v_1) (y^{\alpha}).$$ 
We also have
\bea  {\cal{D}}_0 + {\cal{D}}_1 &=& \partial_0 + \partial_1 +i(v_0 +
v_1), \\
 {\cal{D}}_+ &=&\frac{\partial}{\partial \theta^+} -i {\bar{\theta}}^+
({\cal{D}}_0 + {\cal{D}}_1), \\
 {\bar{\cal{D}}}_+ &=& -\frac{\partial}{\partial {\bar{\theta}}^+} +i
\theta^+  ({\cal{D}}_0 + {\cal{D}}_1), \\
 {\cal{D}}_0 - {\cal{D}}_1 &=& \partial_0 -\partial_1 +iV, \eea
where $V$ is given by
\be V= v_0 - v_1 -2i \theta^+ {\bar{\lambda}}_- -2i {\bar{\theta}}^+ 
\lambda_- +2 \theta^+ {\bar{\theta}}^+ D. \ee
The gauge invariant field strength is $\Upsilon = [{\bar{\cal{D}}}_+
  ,{\cal{D}}_0   - {\cal{D}}_1]$ which has a corresponding action, 
\be S_g = \frac{1}{8 e^2} \int {\rm{d^2}} y \ {\rm{d^2}} \theta \
\bar{\Upsilon} \Upsilon  = \frac{1}{e^2} \int {\rm{d^2}} y \
(\frac{1}{2}v_{01}^2  +i {\bar{\lambda}}_- (\partial_0 +
\partial_1)\lambda_-  +\frac{1}{2} D^2  ). \ee

\subsection{The chiral multiplet}

There are bosonic chiral superfields $\Phi_i^0$ satisfying
${\bar{D}}_+ \Phi_i^0 =0$. Defining ${\Phi_i}^0 = e^{-\Psi} \Phi_i$,
we see that ${\bar{\cal{D}}}_+ \Phi_i =0$. Here $\Phi_i$ has the component expansion
\be \Phi_i = \phi_i +{\sqrt 2} \theta^+ \psi_{+i} -i \theta^+
    {\bar{\theta}}^+ (D_0 + D_1)\phi_i. \ee
This corresponding gauge invariant action is given by
\bea
 S_{ch} &=&  - \frac{i}{2} \int {\rm{d^2}} y \ {\rm{d^2}} \theta \
\sum_{i} {\bar{\Phi}}_i({\cal{D}}_0 -{\cal{D}}_1)\Phi_i \\ &=& \int
    {\rm{d^2}} y \sum_{i} (- \vert D_{\alpha} \phi_i\vert^2 +i
    {\bar{\psi}}_{+i}(D_0 - D_1) \psi_{+i} -i Q_i {\sqrt 2}
    {\bar{\phi}}_i \lambda_- \psi_{+i} \cr & & +i Q_i {\sqrt 2} \phi_i
    {\bar{\psi}}_{+i} {\bar{\lambda}}_-  + Q_i D \vert \phi_i \vert^2)
    \non \eea
where $\Phi_i$ has a $U(1)$ charge $Q_i$.

\subsection{The Fermi multiplet}

There are also fermionic chiral superfields, $\Gamma_a^0$, with
negative chirality  satisfying 
$${\bar{D}}_+ \Gamma_a^0 = {\sqrt 2} E_a^0$$ 
where $E_a^0$ satisfies ${\bar{D}}_+ E_a^0 =0$. Defining $\Gamma_a^0=
e^{-\Psi} \Gamma_a$ and $E_a^0= e^{-\Psi} E_a$, the Fermi superfield has a component expansion
\be \Gamma_a = \chi_{-a} -{\sqrt 2} \theta^+ G_a -i \theta^+
    {\bar{\theta}}^+ (D_0 + D_1) \chi_ {-a} -{\sqrt 2}
    {\bar{\theta}}^+ E_a. \ee
We will consider cases where $E_a=0$. In this case, the kinetic terms
for the Fermi multiplet are given by
\be S_F= -\frac{1}{2} \int {\rm{d^2}} y \ {\rm{d^2}} \theta \ \sum_{a}
    {\bar{\Gamma}}_a \Gamma_a = \int {\rm{d^2}} y \ \sum_{a} (i
    {\bar{\chi}}_{-a}  (D_0 + D_1) \chi_{-a} + \vert G_a \vert^2).\ee

\subsection{The $D \theta$ term}

The terms in the action containing the Fayet-Iliopoulos $D$-term and  
the theta term are given by
\be S_{D \theta}=  \frac{t}{4} \int {\rm{d^2}} y \ {\rm{d}} \theta^+\
\Upsilon  \vert_{\bar{\theta}^+=0} +h.c. = \int {\rm{d^2}} y \ (-r D +
\frac{\theta}{2 \pi } v_{01})\ee
where $t = ir + \frac{\theta}{2 \pi}$.\\

\subsection{The superpotential term}

The $(0,2)$ superpotential is given by
\bea S_J &=& -\frac{1}{\sqrt 2} \  \int {\rm{d^2}} y \ {\rm{d}} \theta^+\  
\sum_{a} \Gamma_a J^a (\Phi_i) \vert_{\bar{\theta}^+=0} -h.c. \cr
&=& - \int {\rm{d^2}} y \ \sum_{a} (G_a J^a (\phi_i) + \sum_{i}
\chi_{-a} 
\psi_{+i} \frac{\partial J^a}{\partial \phi_i} ) -h.c. \eea
where the $J^a$ are functions of the chiral superfields, $\Phi^i$.




\providecommand{\href}[2]{#2}\begingroup\raggedright\endgroup

\end{document}